\documentclass{ws-ijqi}
    \usepackage[utf8]{inputenc}
    \usepackage[english]{babel}
    \usepackage{amssymb,amsmath,graphicx}
    \usepackage{siunitx}
    \usepackage{braket}
    \usepackage{hyperref}
    \usepackage[super,sort,compress]{cite}
    
    \def\rtn{\text{\tiny RTN}}
    \def\ou{\text{\tiny OU}}
    \newcommand{\channel}{\mathcal{E}}
    \DeclareMathOperator{\Tr}{\text{Tr}}
    \begin{document}
    \title{Non-Markovianity by undersampling in quantum optical simulators}
    \author{MATTEO A. C. ROSSI\footnote{matteo.rossi@unimi.it}, CLAUDIA BENEDETTI, DARIO TAMASCELLI}
    \address{Quantum Technology Lab, 
    Dipartimento di Fisica, Universit\`a di Milano, I-20133 Milano, Italy}
    
    \author{SIMONE CIALDI, STEFANO OLIVARES, BASSANO VACCHINI, MATTEO G. A. PARIS\footnote{matteo.paris@fisica.unimi.it} }
    \address{Quantum Technology Lab, 
    Dipartimento di Fisica, Universit\`a di Milano, I-20133 
    Milano, Italy\\ INFN, Sezione di Milano, I-20133 Milano, Italy}
    \maketitle
    \begin{history}
    \received{\today}
    \end{history}
    \catchline{}{}{}{}{}
    \begin{abstract}
    We unveil a novel source of non-Markovianity for the dynamics of quantum
    systems, which appears when the system does not explore the full set of 
    dynamical trajectories in the interaction with its environment. We term 
    this effect {\em non-Markovianity by undersampling} and demonstrate its 
    appearance in the operation of an all-optical quantum simulator involving
    a polarization qubit interacting with a dephasing fluctuating environment.
    \end{abstract}
    \section{Introduction}
    Non-Markovianity is a property of quantum dynamical
    maps which, loosely speaking, should capture the appearance
    of memory effects in the evolution of open quantum systems
    \cite{breu:JPB:12,Rivas2014a,Breuer2016a}.
    Such effects can be traced back to a backflow of information
    from the environment to the system, and their appearance is closely connected to a property of the dynamics known as divisibility
    \cite{wolf:PRL:08,rivas:PRL10,Chruscinski2011a,Wissman2015a,LXM:PRA10,Benatti20122951,fanchini:PRL:14,haseli:PRA:14}. 
    Lack of this property reflects the fact that knowledge of
    the system state at a given time is not enough to
    determine its future evolution.
    \par
    In those situations where the open quantum system is coupled to a
    classical-like fluctuating environment \cite{Zhou2010,AA13,pal14,ben13}, the
    partial trace over the the environment is usually obtained by
    averaging the dynamics over the realizations of the stochastic process that describes its
    classical fluctuations. On the other hand, the possible non-Markovianity
    of the resulting dynamical map is not determined by the sole properties
    of the classical stochastic process. Rather, it results from an interplay
    between the structure of the interaction Hamiltonian, the time scale of
    the classical environment and the dimension of the open quantum system.
    In fact, a classical environment with fluctuations described by a
    (classically) non-Markovian process may induce either a Markovian or a
    non-Markovian quantum evolution, depending on the structure of the
    interaction Hamiltonian \cite{comp14,Rossi2016}.
    \par
    Having these considerations in mind, in this paper we discuss and
    unveil a novel source of non-Markovianity for open quantum systems coupled
    to classical fields, which appears when the structure of the interaction
    does not allow the system to explore the full set of realizations of the
    stochastic process. In this case, the reduced dynamics of the open quantum
    system does not correspond to the averaging over the stochastic ensemble,
    since the system is not actually sensing all the possible trajectories of
    the environment. Rather, the average should be explicitly performed
    on the actual trajectories and the resulting dynamical map may be non-Markovian
    also when the ensemble-averaged one is Markovian.
    \par
    We term this effect
    {\em non-Markovianity by undersampling} and demonstrate its appearance
    in optical platforms, that is, for polarization qubit interacting with a
    dephasing fluctuating environment.  To this aim, we employ our recently
    developed all-optical quantum simulator \cite{Cialdi2017}. In turn, our
    analysis may be considered as a benchmark to assess the performances 
    of quantum simulators involving sampling of limited size. 
    \par
    It is worth noting that for open quantum systems subject
    to dephasing, an effective description in terms of the coupling with
    a classical fluctuating field is always viable 
    \cite{Crow2014,Benedetti2014b,chenu17}. The
    explicit construction of the corresponding classical stochastic
    process may been indeed obtained for a generic quantum environment \cite{chenu17}.
    Non-Markovianity by undersampling is thus expected to be a general
    feature, which is present in any system interacting with a structured
    environment inducing a dephasing dynamics.
    Besides, quantum environments may be described by classical fields,
    at least in the short-time limit, whenever global symmetries are available,
    leading to the definition of environmental operators that remain well
    defined when the size of the environment is increased \cite{Rossi2017b}.
    \par
    The paper is structured as follows: In Sec. \ref{sec:model} we introduce the model of the dynamics, in Sec. \ref{sec:apparatus} we briefly describe the experimental setup and in Sec. \ref{sec:results} we present the experimental results and their analysis. Section \ref{sec:conclusions} concludes with final remarks.
    
    \section{Model}
    \label{sec:model}
    Let us consider a single qubit interacting with a classical
    field via the (interaction) Hamiltonian $H_I= \lambda(t)\, \sigma_3$,
    where $\lambda(t)$ denotes a stochastic process describing the
    fluctuating field and $\sigma_3$ is a Pauli matrix. The corresponding evolution operator is
    given by 
    \begin{equation}
      U(t) = \exp\left\{-i \int_0^t H_I(s)  \, ds\right\} = e^{ - i \varphi(t)\, \sigma_3},
    \end{equation}
    where the time-dependent phase is given by
    $\varphi(t)  = \int_0^t \lambda(s) \, ds$.
    If $\varrho_0$ denotes the initial state of the qubit, the state at time
    $t$ is obtained by averaging over the realizations of the stochastic process,
    that is, the dynamical map corresponds to the ensemble average
    \begin{align}
    \varrho(t) = {\cal E}(t) [\varrho_0]  = \left\langle
    U(t)\varrho_0\, U^\dag(t)\right\rangle_\Lambda\,, \label{eq:exact_map}
    \end{align}
    where the functional integral
    \begin{equation}
    \left\langle f[\lambda(t)]\right\rangle_\Lambda\,
    = \int\! \mathcal{D}[\lambda(t)]\, p[\lambda(t)]\, f[\lambda(t)]
    \end{equation}
    is performed over all the possible trajectories of the stochastic process
    $\Lambda\equiv \lambda(t)$, $p[\lambda(t)]$ being its probability
    distribution and $\mathcal{D}[\lambda(t)]$ being the volume element of 
    the probability space.
    On the other hand, if the interaction between the system and its environment
    is such that the number $N$ of realizations is inherently small, then the
    dynamical map does correspond to the average over the actual realizations,
    i.e.
    \begin{align}\label{eq:undersampled_map}
    \varrho(t) = {\cal E}_N(t) [\varrho_0] = \frac1N \sum_k e^{- i \varphi_k(t)\sigma_3} \varrho_0\,  e^{i \varphi_k(t)\sigma_3} \,,
    \end{align}
    where $\varphi_k(t) = \int_0^t \lambda_k(s) \, ds$ denotes the phase-shift
    originating from the specific $k$-th realization of the process.
    Of course, if the number of realizations is large, we are back to the
    ensemble average by the law of large numbers
    \begin{align}
    {\cal E}_N(t) [\varrho] \stackrel{N \rightarrow \infty}{=}
    {\cal E}_\infty(t) [\varrho] \equiv {\cal E}(t) [\varrho]\,.
    \end{align}
    The properties of the dynamical maps ${\cal E}_N$ at finite $N$ may be
    different from those of ${\cal E}$ and, in particular, ${\cal E}_N$ may
    be non-Markovian even if ${\cal E}$ is Markovian.
    
    We perform simulations involving Gaussian noise and RTN, and
    we discuss the non-Markovianity of the dynamics. For Gaussian noise, we choose a
    paradigmatic example, the Ornstein-Uhlenbeck (OU) stochastic process
    \cite{Uhlenbeck1930}, which has been widely studied in the context of open
    quantum systems \cite{Yu2010,Benedetti2014b,Rossi2016}. Both OU noise and RTN
    are characterized by an exponentially decaying correlation function, and hence
    by a Lorentzian spectrum. The statistics of the two stochastic processes,
    however, is completely different.
    For the RTN, each realization $\lambda_\rtn(t)$ jumps randomly between the two values
    $\pm \nu$, where $\nu$ is a coupling constant, with a switching rate $\gamma$. This means that, after a time $t$, the
    number of jumps that have occurred follows a Poisson distribution with parameter
    $\gamma t$. Thus, in order to generate a sample of RTN noise, we discretize time with
    steps of length $\delta t$ and at each step we perform a jump with probability $\delta
    P= 1-e^{-\gamma \delta t}$. The initial state of the noise is chosen randomly
    between $+\nu$ and $-\nu$, with probability $50 \%$.
    
    \par
    For the OU process, on the other hand, we have
    $\lambda_\ou(t) = \nu B(t)$, where $B(t)$ satisfies the stochastic
    equation 
    \begin{equation}
      B(t + \delta t) = \left( 1-2\gamma \delta
    t \right) B(t) + 2 \sqrt{\gamma} \: dW(t),
    \end{equation}
    where $dW(t)$ is a Wiener increment with zero mean and variance
    $\sigma^2 = \delta t$. For each realization
    we impose the initial condition $B(0)=0$.
    For both models, an exact solution
    for Eq. \eqref{eq:exact_map} can be found \cite{Benedetti2014,Rossi2014}. It reads
    \begin{equation}\label{eq:dephasing_map}
      \rho(t) = \frac{1}{2} [1-G(t)] \sigma_3 \rho_0 \sigma_3 + \frac{1}{2} [1+G(t)] \rho_0.
    \end{equation}
    The function $G(t)$, known as the decoherence function, can be obtained analytically for both noises. For RTN it reads 
    \begin{equation}
      G_{\rtn} = e^{-\gamma t}\left( \cosh{\eta t} + \eta^{-1}
    \gamma \sinh \eta t\right),
    \end{equation}
    where $\eta = \sqrt{\gamma^2 - 4\nu^2}$. For the OU noise:
    \begin{equation}
    G_{\ou} = e^{-2 \nu^2\beta(t)}, \qquad \beta(t) = \frac 1{2 \gamma^2}(e^{-2 \gamma t} +
    2 \gamma t - 1).
    \end{equation} 
    To work with adimensional units, in the following
    we redefine $t$ as $\nu t$ and $\gamma$ as $\gamma / \nu$.
    \par
    Among the different criteria that have been devised to characterize the
    non-Markovianity of a quantum map, we employ the one introduced by Breuer et al.
    \cite{Breuer2009}, which links the presence of a backflow of information from
    the environment to the system to a temporary increase of the distinguishability
    among different initial states of the system evolved according to the same
    reduced dynamics. The distinguishability between states is quantified by their
    trace distance, defined as 
    \begin{equation}
      D(t) = \frac 12 \| \rho_1(t)-\rho_2(t) \|_1,
    \end{equation} 
    where we denote $\|A\|_1 = \Tr \sqrt{A^\dagger A}$ is the trace norm of the operator $A$. A
    map is non-Markovian if there exists a pair of initial states $\rho_1(0),\rho_2(0)$
    for which $D(t)$ is not monotonically decreasing in time.
    
    In Ref. \citeonline{Breuer2009} a measure $\mathcal{N}$ is introduced in order to quantify the degree of non-Markovianity. It is defined as the time integral of the
    derivative of the trace distance on the time intervals where it is increasing,
    that is 
    \begin{equation}\label{eq:blp_definition}
      \mathcal{N}(\channel) = \max_{(\rho_1,\rho_2)}  \int_0^\infty (\dot D_{12}(t) + |\dot D_{12}(t)|)dt,
    \end{equation}
    where 
    \begin{equation}
      \dot D_{12}(t) = \frac{d}{d t} D\left(\rho_1(t),\rho_2(t)\right),
    \end{equation}
    and the maximization is over all possible pairs of initial states of the dynamics.
    $\mathcal{N}(\channel)$ is clearly zero if the map $\channel$ is
    Markovian, and it is greater the more the trace distance deviates for a
    monotonically decreasing behavior.
    \par
    For single-qubit dephasing channels as in Eq. \eqref{eq:dephasing_map}, the
    optimal pair of states to witness non-Markovianity is known
    to be the pair $\ket{\pm} = (\ket{H} \pm \ket{V})/\sqrt{2}$ \cite{Breuer2016a}.
    The trace distance between these two states is
    $D\big(\rho_+(t),\rho_-(t)\big) = | G(t) |$,
    where $\rho_\pm = \ket{\pm}\bra{\pm}$. Thus, a non-monotonic
    behavior of the decoherence function $G(t)$ is a necessary and
    sufficient condition for the non-Markovianity of the channel.
    Starting from he above formula for $D\big(\rho_+(t),\rho_-(t)\big)$,
    it is clear that the dephasing map induced by the Gaussian stochastic
     process is Markovian, as $\beta(t)$ is a
    monotonically increasing function of $t$, while RTN gives a non-Markovian map
    for $\gamma < 2$ \cite{Benedetti2014}. But if the dynamics of the qubit is
    given by a finite number of realizations of the stochastic process, Eq.
    \eqref{eq:undersampled_map}, then the above conclusions are no longer valid.
    \par
    
    \section{Experimental apparatus}
    \label{sec:apparatus}
    In order to demonstrate the non-Markovianity by undersampling, we exploit our recently developed quantum
    simulator \cite{Cialdi2017}. This simulator can perform the evaluation in parallel
    of Eq. \eqref{eq:undersampled_map}
    using the polarization of a single photon as a qubit and exploiting its
    spectral components to average over the realizations of the stochastic
    dynamics. In particular, we consider the qubit affected by dephasing
    driven either by Gaussian noise or non-Gaussian random-telegraph noise (RTN).
    These are interesting examples since, in both cases, the ensemble
    average of Eq. (\ref{eq:exact_map}) may be performed analytically and it is known that
    Gaussian noise is leading to a Markovian map, whereas RTN noise may originate
    both Markovian and non-Markovian maps depending on the values of its switching
    rate \cite{Benedetti2014}.
    \begin{figure}[t]\centering
    \includegraphics[width=0.9\textwidth]{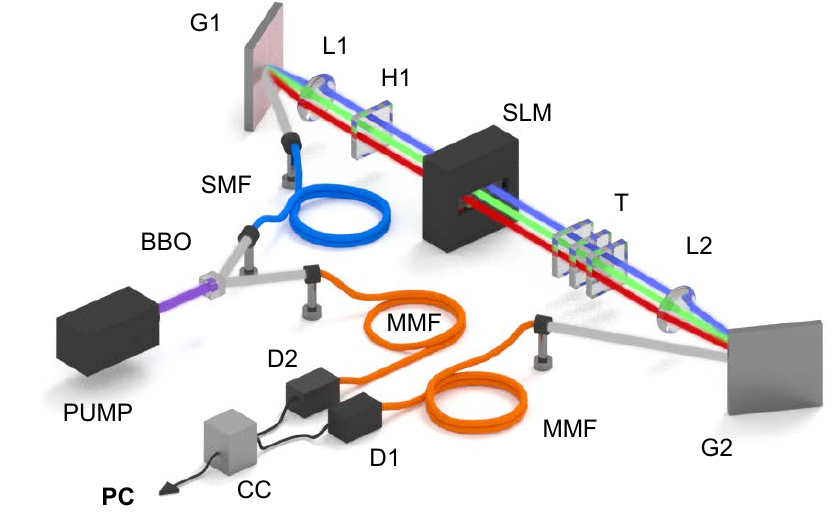}
    \caption{Schematic diagram of our setup. The pump is a
    $\SI{405.5}{\nano\meter}$ laser diode; a couple of frequency-entangled
    photons is generated via parametric down-conversion (PDC) through a BBO,
    Beta barium borate nonlinear crystal; one photon is sent via a
    multi-mode fiber (MMF) to the single-photon detector D2. The other is
    sent through a single-spatial-mode and polarization preserving fiber
    (SMF) to the 4F system. The 4F system is composed of two diffraction
    gratings G1-G2, two lenses L1-L2, a half-wave plate H1 that prepares the
    photon in the initial state $\ket{+}$, the spatial light modulator (SLM),
    and a tomographic apparatus T, made of a quarter-wave plate, a half-wave
    plate and a polarizer.  The photon is then sent through a MMF to the
    single-photon detector D1. Finally, an electronic device measures the
    coincidence counts (CC) and sends them to the computer (PC).}
    \label{fig:apparatus}
    \end{figure}
    \par 
    Our experimental setup is sketched in Fig.~\ref{fig:apparatus} and
    described in detail in Ref. \citeonline{Cialdi2017}. Frequency-entangled photon
    pairs are generated by parametric down-conversion (PDC) and then collected
    by two fiber couplers. The idler photon is detected after traveling
    through a multimode fiber (MMF). The signal photon enters a 4F system and is then coupled to a MMF and reaches the single photon detector. Coincidence counts with the idler
    photon are then detected.
    The key ingredient of the simulator is a spatial light modulator (SLM),
    placed on the Fourier plane between the two lenses L1 and L2 of the 4F
    system. The SLM is a 1D liquid crystal mask ($640$ pixels) used to
    introduce a different phase (externally controlled by the PC)
    to each pixel. The PDC spectrum, selected with a rectangular profile
    through a slit, hits 64 pixels.
    A phase $\varphi_k(t)$ is assigned to each group of pixels, implementing the
    simulation of the dynamical map in Eq. \eqref{eq:undersampled_map}. The average
    over the realizations of the noise is thus performed by (coherently) collecting
    the different spatial components through the lens L2 and the grating G2 into
    a MMF. The state reconstruction is performed by the tomographic
    apparatus T placed between the SLM and the L2 lens.
    
    \section{Results}
    \label{sec:results}
    \begin{figure}[t] 
      \centering
      {\sf \footnotesize 
      \begin{tabular}{ccc} 
      & OU & RTN \\
      \parbox[c]{1em}{\rotatebox{90}{2 realizations}}  & \parbox[c]{5cm}{\includegraphics{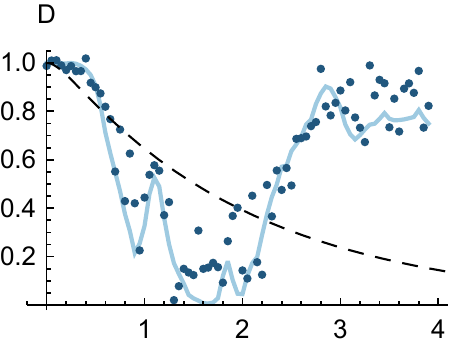}} &
      \parbox[c]{5cm}{\includegraphics{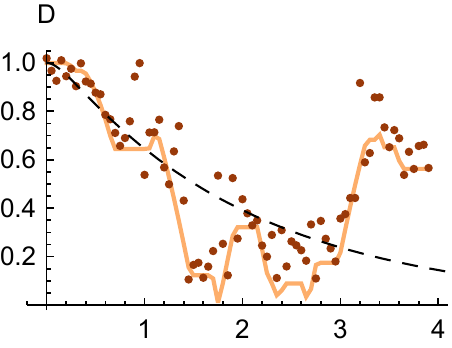}} \\
      \parbox[c]{1em}{\rotatebox{90}{16 realizations}} & \parbox[c]{5cm}{\includegraphics{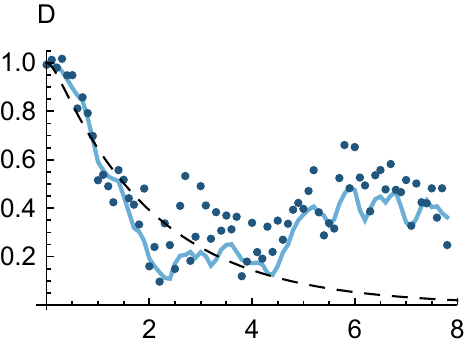}} &
      \parbox[c]{5cm}{\includegraphics{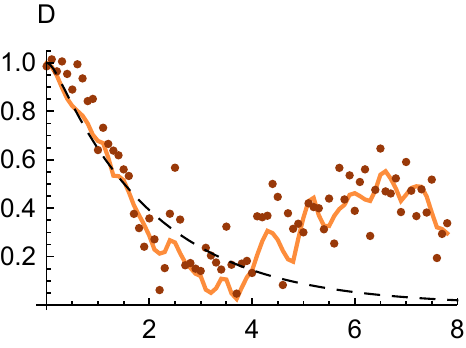}} \\
      \parbox[c]{1em}{\rotatebox{90}{64 realizations}} & \parbox[c]{5cm}{\includegraphics{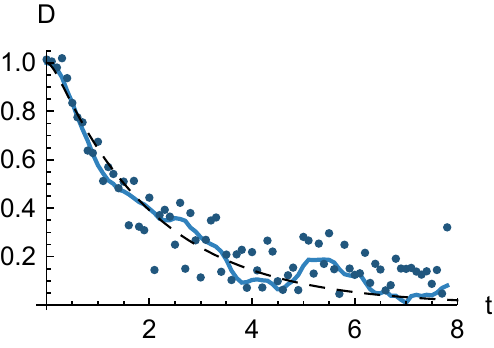}}  &
      \parbox[c]{5cm}{\includegraphics{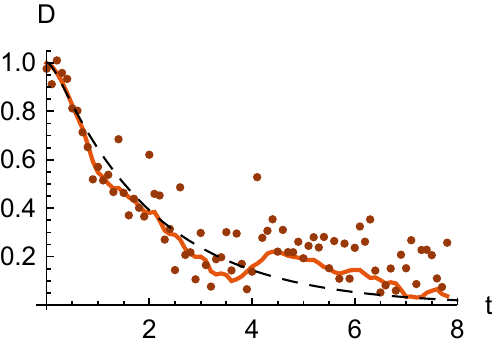}} \end{tabular}}
      \caption{Evolution of the trace distance with time for the OU noise (left) and
      RTN (right) for 2, 16 and 64 realizations of the noise (from top to bottom),
      with $\gamma = 4$. The points represent experimental data, while
      the solid curve is the simulated trajectory. For comparison, the dynamics
      resulting from the ensemble-averaged noise is shown with the black dashed line.
      The trace distance has revivals that are more pronounced for lower numbers of
      realizations. This is a clear signature of the 
      non-Markovianity of the map, in contrast
      with the analytical solution of Eq. \eqref{eq:dephasing_map} that shows a
      monotonic behavior.} \label{fig:td_vs_t}
      \end{figure}
    
    \begin{figure}[t]
      \includegraphics{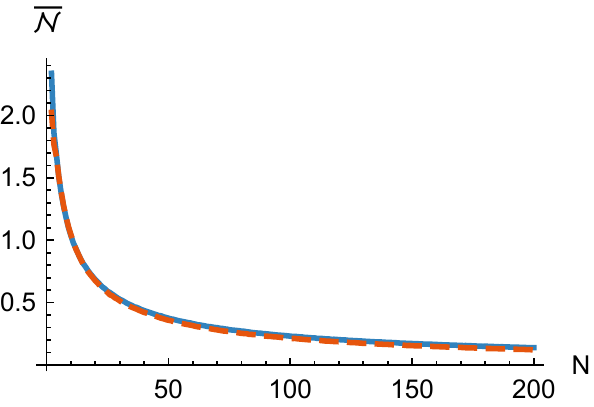}\includegraphics{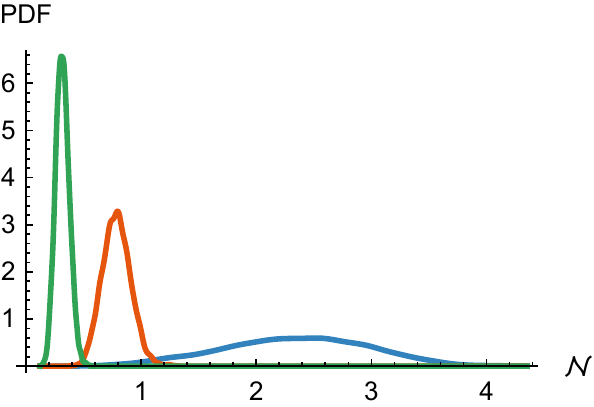}
      \caption{Left panel: Average non-Markovianity $\overline{\mathcal{N}}$ 
      on the time interval
      $t\in[0,8]$, obtained by repeating the simulation 5\,000 times, as a function
      of the number of realizations $N$ of the noise. The solid blue line is for the
       OU noise, the dashed orange line is for the RTN. Right panel: the
       distribution of the values of $\mathcal{N}$ for the OU noise, for the number
       of realizations of Fig. \ref{fig:td_vs_t}: $N = 2$ (blue), $N=16$ (orange),
       $N=64$ (green). }  \label{fig:non_mark_vs_traj}
    \end{figure}
    With the apparatus described above, we simulated the interaction of the qubit,
    initially prepared in the state $\ket{+}$, with RTN and OU noise, with $\gamma =
    4$. For $\gamma = 4$ the RTN is in the fast
    regime and thus for both kinds of noise the dynamics is Markovian when
    considering the ensemble average, Eq. \eqref{eq:exact_map}.
    To perform the simulation, we discretize the time interval $\{0,t_1,...,t_n\}$ (with
    step $\Delta t = 0.001$) and generate on a computer the required number of
    realizations of each type of noise $\lambda_k(t_i)$.
    Then for each output time step $t_i$, the accumulated phases $\varphi_k(t_i) =
    \int_0^{t_i} \lambda_k(t) \, dt$ are encoded in blocks of adjacent pixels in order
    to use the maximum number of available pixels. A photon initially prepared in the
    $\ket{+}$ state is sent through the SLM and its state is
    then reconstructed via a tomography, with four projective measurements \cite{Banaszek1999,James2001,Cialdi:08,Cialdi2014}. The
    acquisition time is $\SI{10}{\second}$. 
    
    From the off-diagonal element of the density matrix
    we can obtain the decoherence function $G(t)$ and hence the optimal trace distance.
    The appropriate corrections are implemented to take into account imperfections in
    the experimental apparatus. The initial state of the photon is not exactly $\rho_+$,
    but rather a combination with the maximally mixed state: 
    \begin{equation}
      \rho_{0,\text{exp}} = p\rho_+ + (1-p) \mathbb{I}/2,
    \end{equation}
    where $\mathbb{I}$ is the identity operator and $p \sim 0.98$.
    \par
    The results are summarized in Fig. \ref{fig:td_vs_t}, which shows the evolution of the trace distance as a function
    of time for the two noises, comparing experimental data (points) with a
    simulation (solid, shaded lines) and with the analytical solution of the
    ensemble-averaged map, Eq. \eqref{eq:exact_map}. From top to bottom, the number of
    realizations of the noise that are simulated in the SLM increases. We clearly
    see that the trace distance has revivals, thus witnessing the non-Markovianity
    of the quantum evolution. The behavior of the map is similar between the two kinds
    of noise. The lower the number of realizations, the more pronounced are the
    revivals. For 64 realizations of the noise, the evolution of the trace distance
    is close to the ensemble-averaged map.
      
    \par
    We now seek to find a relation between the non-Markovianity of the quantum map
    of Eq.~\eqref{eq:undersampled_map} and the number $N$ of trajectories that build
    up the  map. A quantitative analysis must rely on a measure of the degree of
    non-Markovianity of the dynamics. We employ the one introduced in Eq.~\eqref{eq:blp_definition}, based on the backflow of information from the environment to the system, using the pair of optimal initial states $(\ket{+}, \ket{-})$. 
    
    The results are presented in Fig.~\ref{fig:non_mark_vs_traj}. The map $\channel_N(t)$ depends on the actual realizations of the noise and thus we consider the average $\overline{\mathcal{N}}$ of the non-Markovianity measure $\mathcal{N}$ over a large number of repetitions of the experiment. In the left panel, $\overline{\mathcal{N}}$ is presented as a function of the number of realizations $N$ of the noise, for both OU and RTN. We can see that it decreases monotonically with $N$, although the functional dependence is not trivial. On the right panel, the probability density function of $\mathcal{N}$ for different simulation is presented for three different values of $N$, showing that, with increasing $N$, the distribution gets more peaked around the average $\overline{\mathcal{N}}$.
    
    \begin{figure}[t]
      \includegraphics{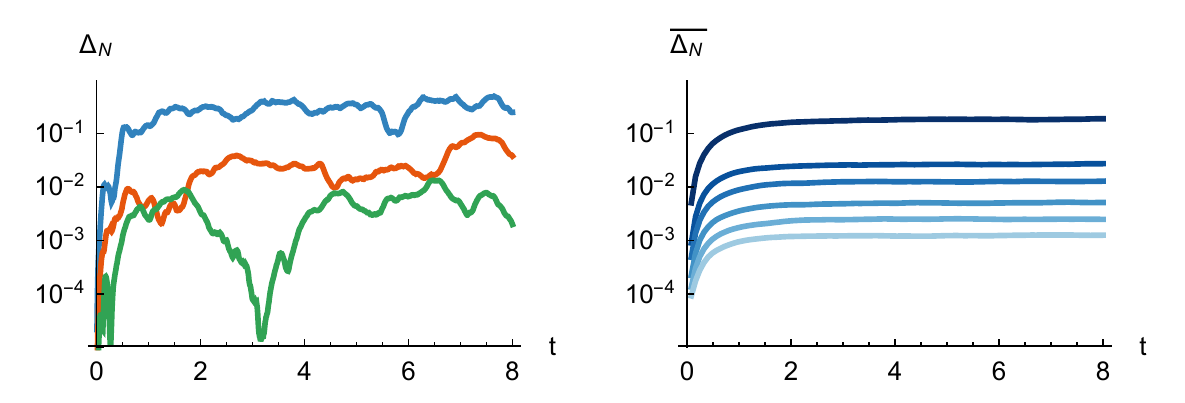}
      \caption{In the left panel, infidelity $\Delta_N$ as a function of
      time for $N=2$ (blue), $N = 16$ (orange), $N = 64$ (green) for OU noise.
      In the right panel the same quantity, averaged over 5\;000 repetitions
      of the experiment, for (top to bottom) 2, 10, 20, 50, 100 and 200 trajectories
      of the noise. We can see that the average fidelity saturates to a constant
      value that depends on $N$.}  
      \label{fig:infidelity_vs_t}
    \end{figure}
    
    From the considerations above a question arises on whether we may link the
    non-Markovianity of the map ${\cal E}_N$ to its {\em distance} from the
    asymptotic one ${\cal E}_\infty$. As we will see, this is indeed the case.
    As a measure of the distinguishability, we employ the
    {\em infidelity} 
    \begin{equation}
    \Delta_N\equiv \Delta_N ({\cal E}_N,{\cal E}_\infty)= 1 -
    {\cal F}({\cal E}_N,{\cal E}_\infty),
    \end{equation}
    where the fidelity
    ${\cal F}({\cal E}_N,{\cal E}_\infty)$
    between channels is defined as the
    state fidelity between the Choi-Jamiołkowski (CJ) states of the two channels \cite{Raginsky2001}.
    Given the maximally entangled state between the qubit and an ancilla,
    $\ket{\Psi} = (\ket{00} + \ket{11})/\sqrt 2$, the CJ state of a map $\channel$
    is $\rho_\channel = (\mathbb{I} \otimes \channel)(\ket{\Psi}\bra{\Psi})$.
    After a straightforward calculation, we then obtain the infidelity between
    the channels $\channel_\infty(t)$ and $\channel_N(t)$ (for the sake of simplicity
    we drop the explicit dependence on $t$ of $G$ and $G_N$):
    \begin{align}
      \Delta_N(t) = &
      \frac{1}{2} \left[1 - G\, \text{Re} [G_N] - \sqrt{\left(G^2-1\right) \left(|G_N|^2-1\right)}\right] \label{eq:fidelity}
    \end{align}
    where $G_N = \braket{e^{-2i\varphi_k}}_N$. Notice that in the limit $N\rightarrow \infty$
    we have $G_N \rightarrow G$ and, thus, $\Delta_N(t) \rightarrow 0$.
    \par
    The map $\channel_N(t)$ depends on the actual realizations of
    the noise and thus we consider the average $\overline{\mathcal{F_N}}(t)$
    of the fidelity over a large number of repetitions of the experiment.
    The left panel of Fig.~\ref{fig:infidelity_vs_t} shows the infidelity
    $\Delta_N$ for particular realizations of the experiment with a
    low number of trajectories of the noise, while the right panel shows the same
    quantity averaged over a high number of repetitions. From the latter, we can see
    that the average infidelity $\overline{\Delta_N}$, starting from zero,
    reaches a value that is constant in time and depends on the number of
    trajectories of the noise.
    
    \begin{figure}[t]
      \includegraphics{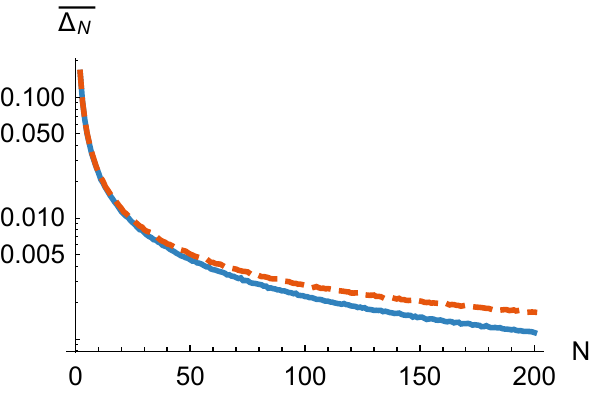} \includegraphics{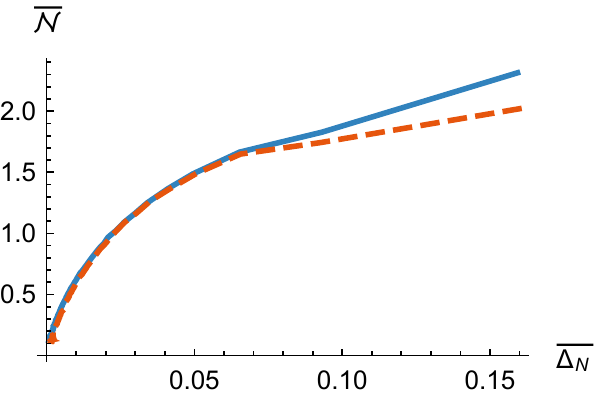}
      \caption{Left panel: Log-log plot of the average over 5\;000 repetitions of the time
      averaged infidelity $ \overline{\Delta_N}$, as a function of the realizations
      of the noise for the OU (solid blue) and RTN (dashed orange). As the number
      of trajectories increases, the infidelity vanishes, as does the BLP measure
      of non-Markovianity. Right panel: Average BLP measure of non-Markovianity $\overline{\mathcal{N}}$ as a function of the
      average infidelity  $ \overline{\Delta_N}$. Notice the monotonic, although non-trivial relation between the two quantities.}
      \label{fig:nonmark_infid}
    \end{figure}
    
    We have then investigated the dependence of this value on $N$ and its
    connection with the non-Markovianity. The results are presented in Fig.~\ref{fig:nonmark_infid}, where the average over time of $\overline{\Delta_N}$ 
    is shown as a function of
    $N$, for the OU noise and RTN, averaged over time. The infidelity decreases
    with $N$ as does the non-Markovianity (cf. Fig.~\ref{fig:non_mark_vs_traj}).
    The right panel clearly shows that there is a
    monotonic dependence of the non-Markovianity measure on the average
    infidelity between the undersampled channel and the ensemble-averaged one.
     
    \section{Conclusions}
    \label{sec:conclusions}
    In this paper we have introduced, demonstrated and discussed {\em non-Markovianity
    by undersampling}, a phenomenon which appears in the dynamics of quantum systems
    interacting with structured environments, when the system does not explore the full set of dynamical trajectories. We have demonstrated experimentally
    its appearance using an all-optical quantum simulator built with a polarization
    qubit interacting with a dephasing fluctuating environment. Our results clearly
    indicate that non-Markovianity is quantitatively linked to the {\it infidelity}
    between the undersampled channel and the ensemble-averaged asymptotic one.
    \par
    Our results pave the way for a deeper understanding of the origin 
    of non-Markovianity in dephasing quantum channels 
    and represent a benchmark to assess the performances 
    of quantum simulators involving sampling of limited size. 
    \section*{Acknowledgements}
    This work has been supported by EU through the collaborative H2020
    project QuProCS (Grant Agreement 641277) and by UniMI through the
    H2020 Transition Grant.
    \bibliographystyle{ws-ijqi}
    \bibliography{us13.bib}

\begin{thebibliography}{10}

\bibitem{breu:JPB:12}
H.-P. Breuer, {\em J. Phys. B} {\bf 45}  (2012)   154001.

\bibitem{Rivas2014a}
A.~Rivas, S.~F. Huelga and M.~B. Plenio, {\em Rep. Prog. Phys.} {\bf 77}
  (2014)   094001.

\bibitem{Breuer2016a}
H.-P. Breuer, E.-M. Laine, J.~Piilo and B.~Vacchini, {\em Rev. Mod. Phys.} {\bf
  88}  (2016)   021002.

\bibitem{wolf:PRL:08}
M.~M. Wolf, J.~Eisert, T.~S. Cubitt and J.~I. Cirac, {\em Phys. Rev. Lett.}
  {\bf 101}  (2008)   150402.

\bibitem{rivas:PRL10}
A.~Rivas, S.~F. Huelga and M.~B. Plenio, {\em Phys. Rev. Lett.} {\bf 105}
  (2010)   050403.

\bibitem{Chruscinski2011a}
D.~Chru{\'s}ci{\'n}ski, A.~Kossakowski and A.~Rivas, {\em Phys. Rev. A} {\bf
  83}  (2011)   052128.

\bibitem{Wissman2015a}
S.~Wi\ss{}mann, H.-P. Breuer and B.~Vacchini, {\em Phys. Rev. A} {\bf 92}
  (2015)   042108.

\bibitem{LXM:PRA10}
X.-M. Lu, X.~Wang and C.~P. Sun, {\em Phys. Rev. A} {\bf 82}  (2010)   042103.

\bibitem{Benatti20122951}
F.~Benatti, R.~Floreanini and S.~Olivares, {\em Phys. Lett. A} {\bf 376}
  (2012) 2951 .

\bibitem{fanchini:PRL:14}
F.~F. Fanchini, G.~Karpat, B.~\ifmmode~\mbox{\c{C}}\else \c{C}\fi{}akmak, L.~K.
  Castelano, G.~H. Aguilar, O.~J. Far\'{\i}as, S.~P. Walborn, P.~H.~S. Ribeiro
  and M.~C. de~Oliveira, {\em Phys. Rev. Lett.} {\bf 112}  (2014)   210402.

\bibitem{haseli:PRA:14}
S.~Haseli, G.~Karpat, S.~Salimi, A.~S. Khorashad, F.~F. Fanchini,
  B.~\ifmmode~\mbox{\c{C}}\else \c{C}\fi{}akmak, G.~H. Aguilar, S.~P. Walborn
  and P.~H.~S. Ribeiro, {\em Phys. Rev. A} {\bf 90}  (2014)   052118.

\bibitem{Zhou2010}
D.~Zhou, A.~Lang and R.~Joynt, {\em Quantum Inf. Process.} {\bf 9}  (2010) 727.

\bibitem{AA13}
A.~D'Arrigo, R.~L. Franco, G.~Benenti, E.~Paladino and G.~Falci, {\em Phys.
  Scripta} {\bf 2013}  (2013)   014014.

\bibitem{pal14}
E.~Paladino, Y.~M. Galperin, G.~Falci and B.~L. Altshuler, {\em Rev. Mod.
  Phys.} {\bf 86}  (2014) 361.

\bibitem{ben13}
C.~Benedetti, F.~Buscemi, P.~Bordone and M.~G.~A. Paris, {\em Phys. Rev. A}
  {\bf 87}  (2013)   052328.

\bibitem{comp14}
C.~Addis, B.~Bylicka, D.~Chru\ifmmode \acute{s}\else
  \'{s}\fi{}ci\ifmmode~\acute{n}\else \'{n}\fi{}ski and S.~Maniscalco, {\em
  Phys. Rev. A} {\bf 90}  (2014)   052103.

\bibitem{Rossi2016}
M.~A.~C. Rossi and M.~G.~A. Paris, {\em J. Chem. Phys.} {\bf 144}  (2016)
  024113.

\bibitem{Cialdi2017}
S.~Cialdi, M.~A.~C. Rossi, C.~Benedetti, B.~Vacchini, D.~Tamascelli,
  S.~Olivares and M.~G.~A. Paris, {\em Appl. Phys. Lett.} {\bf 110}  (2017)
  081107.

\bibitem{Crow2014}
D.~Crow and R.~Joynt, {\em Phys. Rev. A} {\bf 89}  (2014)   042123.

\bibitem{Benedetti2014b}
C.~Benedetti and M.~G.~A. Paris, {\em Int. J. Quantum Inf.} {\bf 12}  (2014)
  ~6.

\bibitem{chenu17}
A.~Chenu, M.~Beau, J.~Cao and A.~del Campo, {\em Phys. Rev. Lett.} {\bf 118}
  (2017)   140403.

\bibitem{Rossi2017b}
M.~A.~C. Rossi, C.~Foti, A.~Cuccoli, J.~Trapani, P.~Verrucchi and M.~G.~A.
  Paris, {\em Phys. Rev. A} {\bf 96}  (2017)   032116.

\bibitem{Uhlenbeck1930}
G.~E. Uhlenbeck and L.~S. Ornstein, {\em Phys. Rev.} {\bf 36}  (1930) 823.

\bibitem{Yu2010}
T.~Yu and J.~H. Eberly, {\em Opt. Comm.} {\bf 283}  (2010) 676.

\bibitem{Benedetti2014}
C.~Benedetti, M.~G.~A. Paris and S.~Maniscalco, {\em Phys. Rev. A} {\bf 89}
  (2014)   012114.

\bibitem{Rossi2014}
M.~A.~C. Rossi, C.~Benedetti and M.~G.~A. Paris, {\em Int. J. Quantum Inf.}
  {\bf 12}  (2014)   1560003.

\bibitem{Breuer2009}
H.-P. Breuer, E.-M. Laine and J.~Piilo, {\em Phys. Rev. Lett.} {\bf 103}
  (2009)   210401.

\bibitem{Banaszek1999}
K.~Banaszek, G.~M. D'Ariano, M.~G.~A. Paris and M.~F. Sacchi, {\em Phys. Rev.
  A} {\bf 61}  (1999)   010304.

\bibitem{James2001}
D.~F.~V. James, P.~G. Kwiat, W.~J. Munro and A.~G. White, {\em Phys. Rev. A}
  {\bf 64}  (2001)   052312.

\bibitem{Cialdi:08}
S.~Cialdi, F.~Castelli, I.~Boscolo and M.~G. Paris, {\em Appl. Opt.} {\bf 47}
  (2008) 1832.

\bibitem{Cialdi2014}
S.~Cialdi, A.~Smirne, M.~G.~A. Paris, S.~Olivares and B.~Vacchini, {\em Phys.
  Rev. A} {\bf 90}  (2014)   050301(R).

\bibitem{Raginsky2001}
M.~Raginsky, {\em Phys. Lett. A} {\bf 290}  (2001) 11 .

\end{thebibliography}
    \end{document}